\documentclass[aps,prb,reprint]{revtex4-2}
\usepackage{graphicx}
\usepackage{color}
\usepackage{amsmath}
\usepackage{amsthm}
\usepackage{amssymb}
\usepackage{physics, ulem}
\usepackage{cases}
\usepackage{mathrsfs}
\usepackage{subfigure}
\usepackage{comment}

\usepackage{hyperref}
\hypersetup{
	citecolor = blue,
	colorlinks = true,
	urlcolor = blue
}

\newcommand{\al}{\alpha}
\newcommand{\e}{\epsilon}
\newcommand{\si}{\sigma}

\newcommand{\p}{\prime}

\newcommand{\sgn}{\mathrm{sgn}}

\newcommand{\lb}{\left(}
\newcommand{\rb}{\right)}
\newcommand{\lB}{\left\{}
\newcommand{\rB}{\right\}}
\newcommand{\LB}{\left[}
\newcommand{\RB}{\right]}

\newcommand{\vv}[1]{\boldsymbol{#1}}
\newcommand{\ca}[1]{\mathcal{#1}}

\begin{document}

\title{Hall effect of ferro/antiferromagnetic wallpaper fermions}

\author{Koki Mizuno}
\affiliation{Department of Physics, Nagoya University, Nagoya 464-8602, Japan}
\author{Ai Yamakage}
\affiliation{Department of Physics, Nagoya University, Nagoya 464-8602, Japan}

\date{\today}

\begin{abstract}
Nonsymmorphic crystals can host characteristic double surface Dirac cones with fourfold degeneracy on the Dirac points, called wallpaper fermion, protected by wallpaper group symmetry.
We clarify the charge and spin Hall effect of wallpaper fermions in the presence of the (anti)ferromagnetism.
Based on a four-sublattice model, we construct the effective Hamiltonian of wallpaper fermions coupled with the ferromagnetic or antiferromagnetic moment.
Both ferromagnetic and antiferromagnetic moments induce an energy gap for the wallpaper fermions, leading to quantized (spin) Hall conductivity. 
The ferromagnetic wallpaper fermion induces the Hall conductivity quantized into $e^2/h$, which is twice that for a single Dirac fermion on the surface of topological insulators.
On the other hand, the spin Hall conductivity decays and reaches to be a finite value as the antiferromagnetic coupling increases. 
We also show that the results above are valid for a general model of wallpaper fermions from symmetry consideration. 
\end{abstract}

\maketitle

\section{Introduction}
Topological insulators (TIs) host a single Dirac fermion on their surfaces, gapless surface excitations protected by time-reversal symmetry (TRS) \cite{Kane_Mele_TO, Kane_Mele_graphene, Fu_Kane_3DTI, Moore_Malents, Roy2009-os, Hansen_Kane, Qi_Liang_Zhang, Tanaka_Sato_Nagaosa, Ando_topo}.
In recent years, there has been growing interest in characteristic topological surface states that arise from nonsymmorphic space-group symmetry \cite{Shiozaki_nonsymmorhic_TCI, Wang2016-uc, Liu_nonsymmorphic_TCI, Bernevig_KHgSb_nonsymmorphic,Kruthoff_band}.
One is the surface states protected by two glides, called wallpaper fermion \cite{wieder2018wallpaper, glide_high_order_TI_2021, Kondo_insulator_PuB4}, and has a fourfold-degenerate Dirac point accompanied by linear dispersions. 
Therefore, wallpaper fermion potentially exhibits phenomena distinct from those in TIs, and can be a platform for a highly efficient device with novel functionality, such as spintronics. 

A single Dirac fermion on the surface of TIs yields unique spintronic properties, thanks to its topological and spin-momentum-rocking natures \cite{He2019-ax, Tokura2019-cs, He2022-qx}. 
On the other hand, one promising direction of spintronics is to use an antiferromagnet (AFM) \cite{SHE_AFM, Laerge_AHE_AFM, SOC_AFM, AFM_spintronics}, owing to their varieties and high-frequency response \cite{AFM_terahertz_2011,nishitani2010terahertz,kimel2004laser,Ultrafast_AFM_Duong,kimel2005ultrafast}. 
Differently from a single Dirac fermion on the surface of TIs, a wallpaper fermion can be coupled with both ferromagnetic (FM) and AFM moments, due to the fourfold degeneracy stemming from the spin and sublattice degrees of freedom.
Therefore, a wallpaper fermion is a candidate for the AFM spintronics with topological electronic states \cite{Smejkal2018-vb, Feng2019-fo}. 

In this study, we clarify the fundamental properties of the FM and AFM wallpaper fermions, deriving an effective model from a four-sublattice model.  
Both FM and AFM couplings are shown to induce an energy gap for the wallpaper fermions. 
The Hall conductivity for the FM case is quantized into twice the value predicted for conventional TIs.
The spin Hall conductivity (SHC) decreases with increasing the FM/AFM coupling.
In the strong coupling limit, the SHC vanishes for the FM case, while it remains finite value for the AFM case. 
The model is proven to be generic for wallpaper fermions from symmetry consideration. 

This paper is organized as follows.
In Sec.~\ref{2}, we derive an effective Hamiltonian of wallpaper fermion.
Next, we consider the effect of FM and AFM coupling in Sec.~\ref{MWPF}.
In this section, we show the behavior of Hall conductivity, SHC, and degeneracy of eigenvalues of Hamiltonian.
In Sec.~\ref{more_G_framework}, we prove that the effective Hamiltonian is sufficiently generic from the viewpoint of symmetry. 
We give some comment about our model in Sec.~\ref{Discuss}, and summarize our work in Sec.~\ref{summary}.

\section{Effective model for surface states}
\label{2}
\begin{figure}
  \centering
  \includegraphics[scale=0.3]{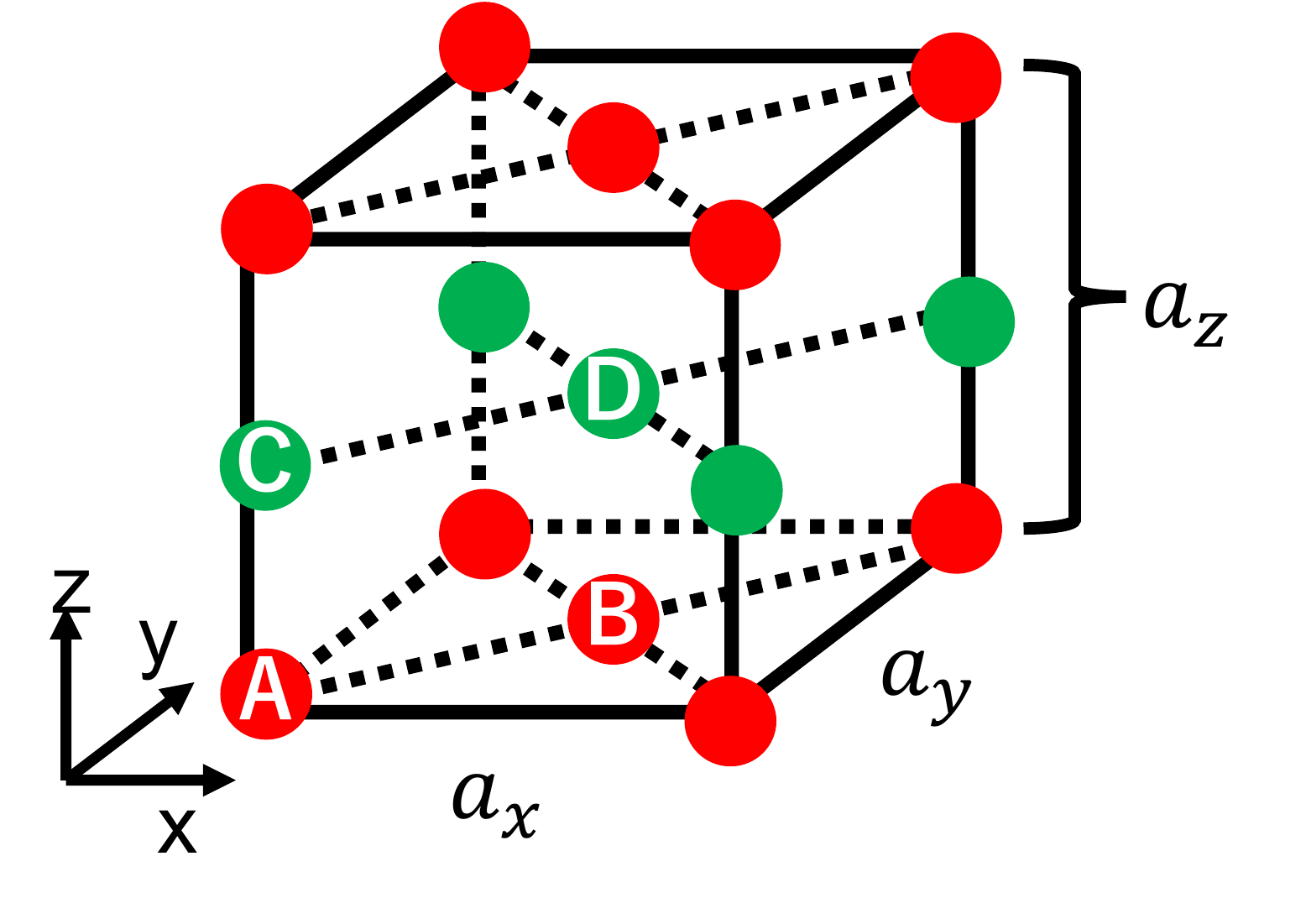}
  \caption{Crystal structure of the toy model. 
  	This crystal has symmetry of SG $P4/mbm$ (No.~127) ($D_{4h}$).
  The model also has mirror symmetry in the $z=1/4,3/4$ plane.
  }
  \label{SG_55}
\end{figure}

In this section, we construct an effective Hamiltonian of the wallpaper fermion on the $xy$ surface in a four-sublattice model \cite{wieder2018wallpaper} with space group $P4/mbm$ (No.~127), as depicted in Fig.~\ref{SG_55}.
The obtained Hamiltonian is beneficial for calculating physical quantities such as (spin) Hall conductivity, as discussed in the subsequent sections. 

\subsection{Bulk Hamiltonian}
\label{bulk}

First of all, we review a four-sublattice model given in Ref.~\cite{wieder2018wallpaper}. 
The crystal structure of the model is the square lattice stacking along the $z$ axis, and consists of four sublattices A, B, C, and D. 
The A and B (C and D) sublattices are located on the $z=0$ ($z=1/2$) plane. 

The Hamiltonian $\mathcal H(\vb*{k}) = \mathcal H_{xy}^1(\vb*{k}) + \mathcal H_{xy}^2(\vb*{k}) + V_z(\vb*{k})$ for bulk is given by
\begin{align}
    \ca{H}^{1}_{xy}(\vb*{k})
    &=\cos\lb\frac{k_xa}{2}\rb
      \cos\lb\frac{k_ya}{2}\rb
      \LB t_1\tau_x+v_{r1}\tau_y\si_z \RB 
      \notag\\& \quad
      +\sin\lb\frac{k_xa}{2}\rb\cos\lb\frac{k_ya}{2}\rb
       \LB v_{s1}\tau_x\mu_z\si_y \RB 
       \notag\\ & \quad
       -\cos\lb\frac{k_xa}{2}\rb\sin\lb\frac{k_ya}{2}\rb
       \LB v_{s1}\tau_x\mu_z\si_x \RB,
       \label{Ham_inpane_1}
 \\
    \ca{H}^{2}_{xy}(\vb*{k})
    &= t_{2} \qty[\cos(k_xa)+\cos(k_ya)] 
    \notag\\&\quad
    + \sin(k_xa)
       \qty [v_{s2}\tau_z\mu_z\si_x
          +v_{s2}^\p\mu_z\si_y ] 
          \notag\\
    & \quad 
    + \sin(k_ya)
       \qty[ v_{s2}\tau_z\mu_z\si_y
          -v_{s2}' \mu_z\si_x ],
          \label{Ham_inpane_2}
\\
    V_z(\vb*{k})
    &= \cos\lb\frac{k_z a_z}{2}\rb u_1\mu_x
       +\sin\lb\frac{k_z a_z}{2}\rb u_2\mu_y 
       \notag\\
        &\quad
        + \cos\lb\frac{k_z a_z}{2}\rb
         \lB
           \cos\lb k_xa \rb
            +\cos\lb k_ya \rb
         \rB
         v_1\mu_x 
         \notag\\
      &\quad +\sin\lb\frac{k_z a_z}{2}\rb
         \lB
           \cos\lb k_xa \rb
            +\cos\lb k_ya \rb
         \rB
         v_2\mu_y,
\end{align}
where $\tau$, $\mu$, and $\sigma$ are the Pauli matrices acting in the inplane sublattice (A,B) [(C, D)], layer sublattice (A, C) [(B, D)], and spin spaces, respectively. 
$\ca{H}_{xy}^1$ is the nearest-neighbor hopping in the same layer,
$\ca{H}_{xy}^2$ is the next-nearest-neighbor hopping in the same layer, and $V_z$ is the spin-independent hopping between neighboring layers.

\subsection{Surface state on the \texorpdfstring{$\bar M$}{M} Point}
\label{surface_state}
Next, we obtain the effective Hamiltonian of the surface states near the $\bar M$ point ($k_x a = k_y a=\pi$), where the wallpaper fermions appear.
To derive the effective Hamiltonian, we first obtain the wavefunction of the surface state for $k_x a = k_y a = \pi$. 
From this solution, we can construct an effective theory in the vicinity of the $\bar M$ point by perturbation expansion.

The Hamiltonian for $k_x a = k_y a = \pi$ in the bulk, $\mathcal H_{\bar M} \equiv \mathcal H(\pi/a,\pi/a,k_z)$, is written as
\begin{align}
    \ca{H}_{\bar M}
    = -2t_{2}
      +A_1\mu_x\cos\lb\frac{k_z a_z}{2}\rb 
       +A_2\mu_y\sin\lb\frac{k_z a_z}{2}\rb,
\end{align}
with $A_1=u_1-2v_1$ and $A_2=u_2-2v_2$.
We further transform the above Hamiltonian by $U= e^{i k_z a_z \mu_z /4}$ as 
\begin{align}
 H_{\mathrm{SSH}}(k_z) \equiv U^\dag \mathcal H_{\bar M} U
 = -2t_2 + \vb*{R} \cdot \vb*{\mu},\label{Ham_SSH}
\end{align}
with
\begin{align}
	R_x &= \frac{A_1-A_2}{2} + \frac{A_1+A_2}{2} \cos(k_za_z) ,
	\\
	R_y &= 
	\frac{A_1+A_2}{2} \sin(k_z a_z),
	\\
	R_z &= 0.
\end{align}
The above is equivalent to the Su-Schrieffer-Heeger (SSH) model \cite{SSH},
which is of class BDI \cite{Ryu_2010}, and has the $\mathbb Z$ classification by the winding number, as shown below.

We can define the winding number $\nu_{w} \in \mathbb Z$ of the parameter vector $(R_x, R_y)$ wrapping the origin by
\begin{equation}
    \begin{split}
        \nu_{w} &\equiv \frac{i}{2\pi}\int^{\pi}_{-\pi} dk_z q^{*}\pdv{q}{k_z},
        \quad
        q = \frac{R_x - i R_y}{\sqrt{R_x^2+R_y^2}},
    \end{split}
\end{equation}
which corresponds to the number of surface zero modes. 
Therefore, when $A_1A_2 < 0$, the winding number satisfies $\nu_{w}=0$.
On the other hand, when $A_1A_2 > 0$, the winding number satisfies $\nu_{w}=1$.
From above, we can find that this Hamiltonian is topologically nontrivial when $A_1A_2>0$.

To obtain the concrete form of wavefunction for the surface state, we solve the tight-binding model $H_{\rm SSH}$ in the semi-infinite ($z \leq 0$) space as
\begin{equation}
    \begin{split}
        H_{\rm SSH}
        &= \sum_{\nu, \rho} 
        \sum_{i=-\infty}^0
        \LB c^{\dag}_{\nu,i}\e_{\nu,\rho}c_{\rho,i} +
        \qty( c^{\dag}_{\nu,i-1}t_{\nu,\rho}c_{\rho,i} +h.c.) \RB, \\
        \e &= \mqty(-2t_2 & \displaystyle\frac{A_1-A_2}{2} \\ \displaystyle\frac{A_1-A_2}{2} & -2t_2), 
        \quad
        t = \mqty(0 & 0 \\ \displaystyle\frac{A_1+A_2}{2} & 0),
    \end{split}
\end{equation}
where $c_{\nu, i}^\dag$ is the creation operator of the fermion on  the $\nu=\mathrm{AB}, \mathrm{CD}$ layer at the $i$th site. 
The system is terminated at $i=0$ by the CD layer. 
The Schr\"{o}dinger equation is given by $H_{\rm SSH}\ket{\psi} = E \ket{\psi}$. The wavefunction has the form 
\begin{equation}
  \begin{split}
      \ket{\psi} &= \sum_{\nu} 
      \sum_{i=-\infty}^0
      \al_{\nu,i}c^{\dag}_{\nu,i}\ket{0},
  \end{split}
\end{equation}
where $\ket{0}$ is the vacuum which satisfies $c_{\nu, i}\ket{0} = 0$.
From the above Schr\"{o}dinger equation, we can find the recurrence form of $\alpha_i = (\alpha_{\mathrm{AB}, i}, \alpha_{\mathrm{CD}, i})^{\mathrm{T}}$ as follows;
\begin{align}
	&
        \e \al_{i} + t\al_{i+1} + t^{\dag}\al_{i-1} = -2t_2 \al_{i}, 
        \qfor i \leq -1,
                \label{recu_formula}
        \\&
        \epsilon \alpha_0 + t^\dag \alpha_{-1} = -2 t_2 \alpha_0,
        \label{BC}
\end{align}
for the state with the energy $E = -2t_2$. 
Now, we assume an exponential form $\al_{i} = \lambda^i \vv{u}$, where $\vv{u} = (u, v)^{\mathrm{T}}$ is a 2-spinor.
Equation (\ref{recu_formula}) is rewritten as follows
\begin{align}
\qty[ {A_1-A_2}+ \qty({A_1+A_2}) \lambda^{-1} ] v = 0,
\\
\qty[ {A_1-A_2} + \qty({A_1+A_2}) \lambda ] u = 0,
\label{eq_Sch_SSH}
\end{align}
The solutions are obtained to be
\begin{equation}
    \lambda = -\frac{A_1+A_2}{A_1-A_2},
    \
    \vb*{u} = \pmqty{
  0
  \\
  1   
 },
\end{equation}
and
\begin{align}
    \lambda = -\frac{A_1-A_2}{A_1+A_2},
 \
 \vb*{u} = \pmqty{
 	1
 	\\
 	0   
 }.
\label{sol2}
\end{align}
For $A_1 A_2 > 0$, the former solution decays into $i \to -\infty$, $|\lambda| > 1$.
The normalized form is given by
\begin{align}
 \ket{\psi} = \frac{2 \sqrt{A_1A_2}}{A_1+A_2}
 \sum_{i=-\infty}^0 
 \qty(- \frac{A_1+A_2}{A_1-A_2})^i c_{\mathrm{CD}, i}^\dag \ket{0},
\end{align}
which also satisfies the boundary condition Eq.~(\ref{BC}). 
On the other hand, for $A_1 A_2 < 0$, Eq.~(\ref{sol2}) is a decaying function but does not satisfy Eq.~(\ref{BC}). 
This means that the zero-energy surface state appears only for $A_1 A_2 > 0$, which is consistent with the discussion based on the winding number. 
Note that the zero-energy surface states in the whole system are fourfold degenerate with respect to $\tau$ and $\sigma$ degrees of freedom.

Finally, we derive the effective Hamiltonian of the wallpaper fermion
by mapping the bulk Hamiltonian onto the surface states at the $\bar M$ point.
The surface states derived above consists only $c^\dag_{\mathrm{CD}, i}$ component, where all the states have $\mu_z = -1$ then $\mu_x = \mu_y = 0$.
Therefore, only the inplane hoppings (\ref{Ham_inpane_1}) and (\ref{Ham_inpane_2}) appear in the effective model.
We obtain the effective Hamiltonian 
\begin{align}
 H_{\mathrm{wp}} = \left. \mathcal{H}_{\bar M} \right|_{\mu_z = -1, \mu_x = \mu_y = 0}  = H_{\mathrm{wp}}^1 + H_{\mathrm{wp}}^2,
 \label{Hsurface}
\end{align}
 with
\begin{align}
	H_{\mathrm{wp}}^1
	&= 
  {-}\frac{v_{s1}}{2}\tau_x(k_x\si_x - k_y\si_y) 
    +\frac{v_{s2}}{2}\tau_z(k_x\si_x + k_y\si_y) 
    \notag\\
    &\quad
    +\frac{v_{s2}^{\p}}{2}\tau_0(k_x\si_y - k_y\si_x) 
    \notag\\
  H_{\mathrm{wp}}^2 &=
    \frac{1}{4}(t_1\tau_{x}+v_{r1}\tau_{y}\si_{z})k_{x}k_{y},
    \label{WP_eff1}
\end{align}
up to the second order of $k$, where $k_x$ and $k_y$ are measured from the $\bar M$ point.
Here we set $a=1$ for simplicity. 
The last term, proportional to $k_x k_y$, is derived from the first-order expansion of Eq.~(\ref{Ham_inpane_1}), and the second-order expansion of Eq.~(\ref{Ham_inpane_2}) is neglected, assuming the band gap is sufficiently large. 
In addition, the term $t_2 (k_x^2+k_y^2) \sigma_0 \tau_0$ is also omitted in order to reproduce only double Fermi surfaces (lines) in the Brillouin zone projected onto the ($xy$) surface. 
This assumption can be justified by introducing the cutoff momentum or higher-order terms.

\begin{figure}
  \centering
  \subfigure[3d plot]{
    \includegraphics[scale=0.4]{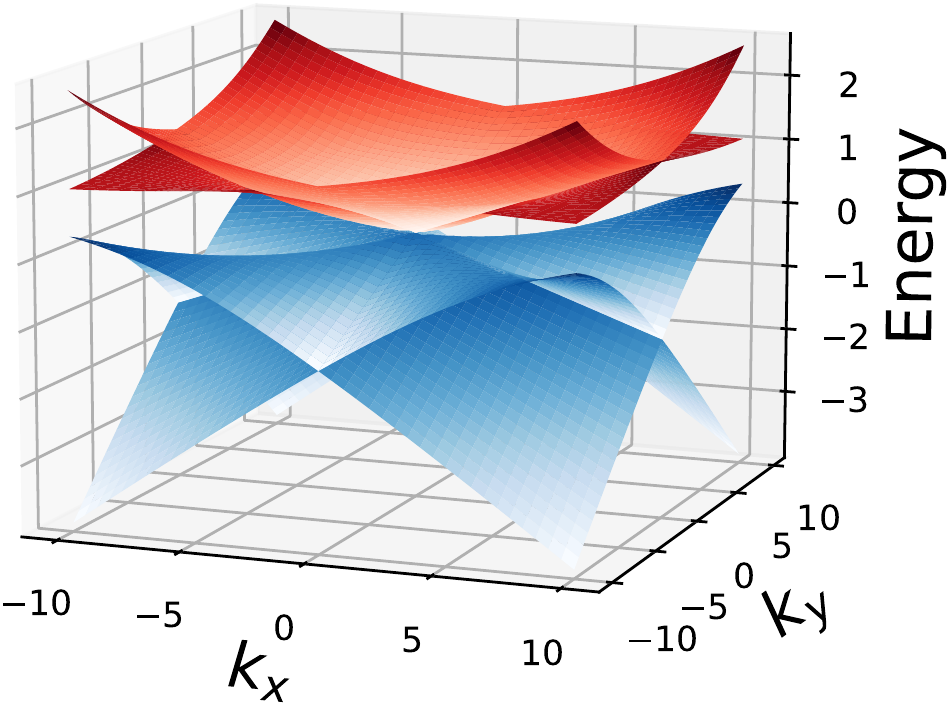}}
  \subfigure[2d plot]{
    \includegraphics[scale=0.65]{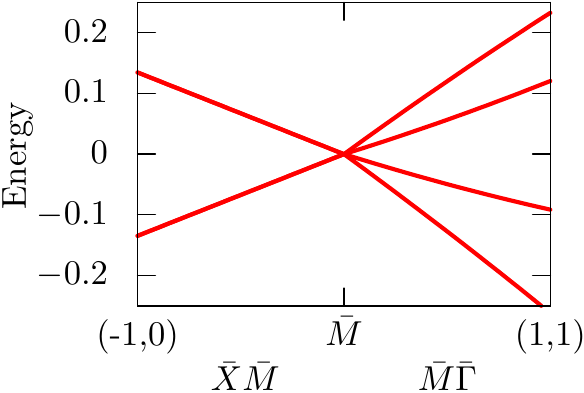}}
  \caption{Energy dispersion of the wallpaper fermion. The parameters are taken as $t_1=0.03$,
  $v_{r1}=0.05$,
  $v_{s1}=0.1$,
  $v_{s2}=-0.2$,
  and
  $v_{s2}^{\p}=0.15$.
  (a) Gapless dispersion around $\bar{M}$ point. There is fourfold degeneracy for $k_x=k_y=0$.
  (b) The corresponding 2d plot along the $k_y=0$ ($\bar X \bar M$) and $k_x = k_y$ ($\bar M \bar\Gamma$) lines. 
  Each dispersion on the $\bar X \bar M$ line is twofold degenerate due to the glide symmetry of the system.}
  \label{Ene_nogap}
\end{figure}

The energy dispersion is shown in Fig.~\ref{Ene_nogap}.
We can find the fourfold-degenerate point protected by the double glide and time-reversal symmetries (TRS) for $k_x=k_y=0$ and twofold-degenerate dispersion protected only by the glide on the $k_y=0$ line.

\subsection{Effective linear model}
\label{trans_effecive_hamiltonian}
This subsection shows that the linear terms $H_{\rm wp}^1$ of the effective Hamiltonian are equivalent to two independent Dirac fermions and that the quadratic terms $H_{\rm wp}^2$ give a mass term and hybridize the Dirac fermions.

We define the unitary matrix $V = e^{-i\phi\tau_z \si_z/2}$ where $\phi = \arg(v_{s2}+iv_{s2}')$.
The effective Hamiltonian is transformed by this matrix as
\begin{equation}
  \begin{split}
    H_{\rm wp}^{1\p}
    &= V^{\dag}H_{\rm wp}^{1} V \\
    &=\frac{v_2}{2}\tau_z(k_x\si_x + k_y\si_y) 
    - \frac{v_{s1}}{2}\tau_x(k_x\si_x-k_y\si_y),
  \end{split}
\end{equation}
where $v_2 = \sqrt{v_{s2}^2+{v_{s2}^{\p 2}}}$.
The effective theory has a conserved charge $X$;
\begin{equation}
  X = \tau_y \si_z,
  \
  \qty[X, H_{\mathrm{wp}}^{1'}] = 0.
\end{equation}
Here we apply the following unitary transform
\begin{align}
	&
 P = \frac{1}{\sqrt 2}
 \qty(
  \ket{+}_{\tau} \ket{+}_\sigma,
  \ket{-}_{\tau} \ket{-}_\sigma,
  \ket{-}_{\tau} \ket{+}_\sigma,
  \ket{+}_{\tau} \ket{-}_\sigma
 ),
 \\&
 \tau_y \sigma_z \ket{\tau}_\tau 
 \ket{\sigma}_\sigma
 = \tau \sigma 
 \ket{\tau}_{\tau} \ket{\sigma}_\sigma,
\end{align}
as
\begin{align}
    P^\dag X P = \mathrm{diag}(1,1,-1,-1).
\end{align}
Therefore, using this matrix $P$, the Hamiltonian is decomposed into $H_{\mathrm{wp}}^\pm$ in the $X = \pm 1$ sectors as
\begin{align}
 H_{\rm wp}^{1^{\p\p}}
 =P^\dag H_{\rm wp}^{1^\p}P
 = \pmqty{
   H_{\rm wp}^+ & 0
   \\
   0 & H_{\rm wp}^-
 }, \label{Ham_first_block}
\end{align}
with
\begin{align}
	H_{\rm wp}^{\pm}
	= \frac{v_2}{2}(k_x\si_x + k_y\si_y)
	\mp
	\frac{v_{s1}}{2}(k_x\si_y + k_y\si_x).
	\label{Ham_wp_pm}
\end{align}
This $2 \times 2$ representation is useful for evaluating the Hall conductivity in the ferromagnetic case, as discussed in the subsequent section.
On the other hand, the quadratic part $H_{\mathrm{wp}}^2$ hybridizes the $X = \pm 1$ sectors.
The second-order term $H_{\mathrm{wp}}^2$ is transformed as
\begin{align}
    H_{\rm wp}^{2^{\p\p}}
    &= P^\dag V^\dag H_{\mathrm{wp}}^2 V P 
    \notag\\
    &=\frac{T_{v}}{4} 
    \qty[\sin(\delta+\phi) \tau_{y}\si_{z} + \cos(\delta+\phi) \tau_{z}\sigma_{0} ] k_{x}k_{y}, \label{trans_H2}
\end{align}
where we define as $v_{r1}+it_{1}=T_{v}\exp(i\delta)$.

\section{Magnetic Wallpaper Fermion}
\label{MWPF}
\begin{figure}
  \centering
  \subfigure[FM coupling]{
    \includegraphics[scale=0.2]{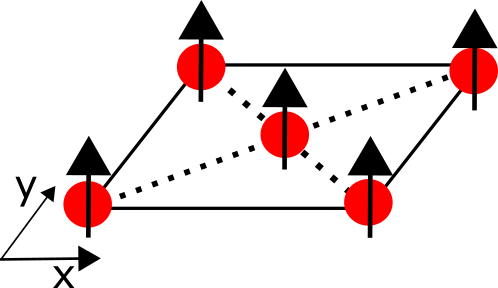}}
  \subfigure[AFM coupling]{
    \includegraphics[scale=0.2]{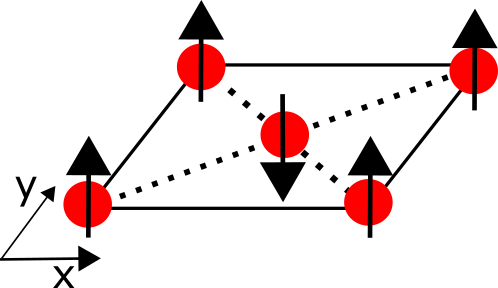}}
  \caption{Schematic of the FM (a) and AFM (b) moments perpendicular to the $xy$ surface.
  }
  \label{fig_magnet}
\end{figure}

In this section, we consider the wallpaper fermions coupled with FM or AFM moment, as shown in Fig.~\ref{fig_magnet}.
In general, the gap opening in a gapless topological state can trigger a transition to a distinct topological phase \cite{Topo_corre_2021, Haruki_Watanabe_mag}.
A representative example of the transition is the surface of 3D TIs; a single Dirac fermion on them can be coupled with only FM moment in the long wavelength limit $k \simeq 0$ and opens the energy gap. 
In contrast, a wallpaper fermion can be coupled with both FM and AFM moments, owing to double degrees of freedom originating from a nonsymmorphic sublattice structure. The resulting charge and spin Hall effects are discussed as below. 

\subsection{Ferromagnetic Wallpaper Fermion}
\label{FM_WPF}

Ferromagnetic moments induce an energy gap in the wallpaper fermion, resulting in the quantization of Hall conductivity, as in a single Dirac fermion on the surface of a topological insulator. 
In this subsection, we verify that the Hall conductivity is quantized into double that in a single Dirac fermion.
Furthermore, we point out that the linear model derived in the previous section exhibits a singular behavior, the plateau transition without gap closing.
On the other hand, the plateau transition disappears in the model, including the second-order terms.
These results imply that the linear model, often used in Dirac systems, is inappropriate for the wallpaper fermions.

\begin{figure}
	\centering
	\subfigure[3d plot]{
		\includegraphics[scale=0.4]{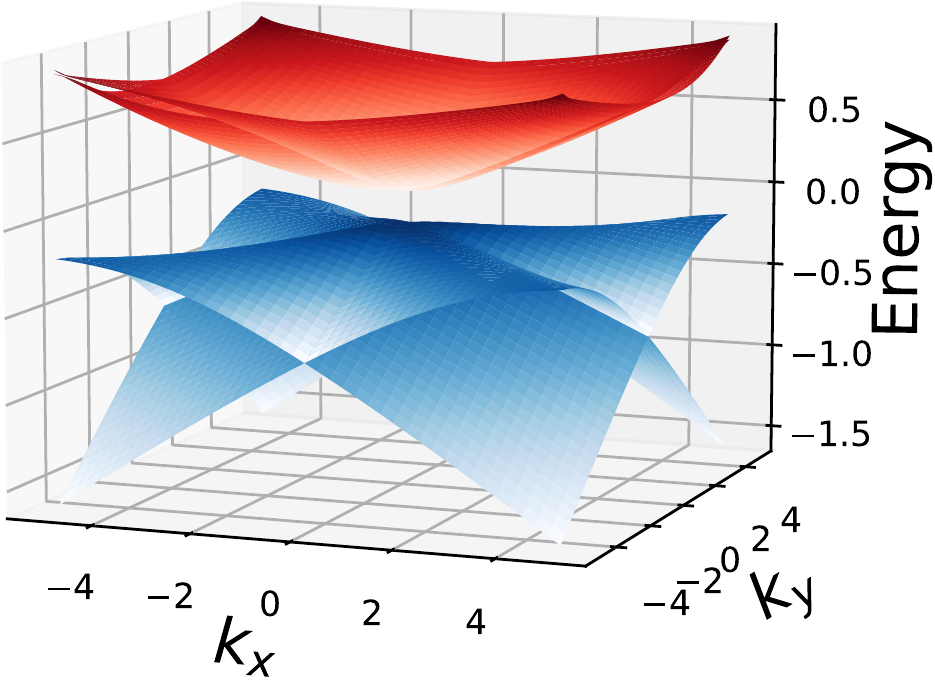}}
	\subfigure[2d plot]{
		\includegraphics[scale=0.65]{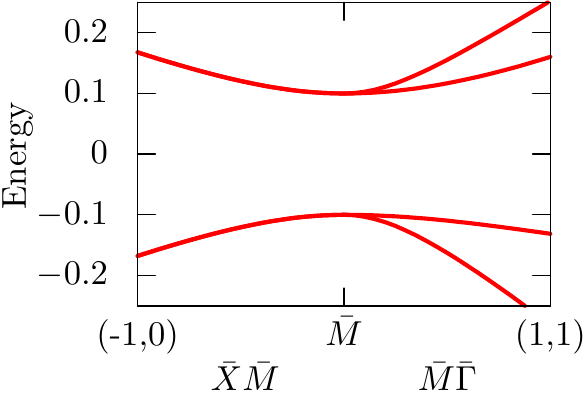}}
	\caption{
		Energy dispersion in the presence of the FM coupling around the $\bar M$ point with 
		$t_1=0.03, v_{r1}=0.05, v_{s1}=0.1, v_{s2}=-0.2, v_{s2}^{\p}=0.15$, and $M=0.1$.
		The twofold degeneracy on the line $k_y=0$ ($\bar X \bar M$ line) remains.}
	\label{Energy_FM}
\end{figure}

First, we consider the FM case as Fig.~\ref{fig_magnet}(a).
The Hamiltonian of the FM coupling is written as
\begin{equation}
  H_{\rm F} = M\si_{z}\tau_{0}, 
  \label{Ham_FM}
\end{equation}
in the original basis adapted in Eq.~(\ref{WP_eff1}).
This Hamiltonian $H_{\rm F}$ is decomposed into the $X=\pm 1$ sectors
\begin{equation}
  H_{\rm F}^{\pm} = M\si_{z}.
\end{equation}
The eigenvalues of the Hamiltonian $H_{\rm wp}^X + H_{\rm F}^X$ are given by $\pm E_X(\vb*{k})$ with
\begin{align}
	E_\pm(\vb*{k})
	= \sqrt{\frac{v_2^2 + v_{s1}^2}{4} k^2 \mp v_2 v_{s1} k_x k_y + M^2},
 \label{Epm}
\end{align}
and are shown in Fig.~\ref{Energy_FM}.
The FM moment $M$ induces an energy gap.

Next, we show the Hall conductivity from the linear response theory.
The Hall conductivity is written as
\begin{align}
    \si_{xy} &=
    -{i\hbar e^2}
     \int \frac{d^2k}{(2\pi)^2} 
     \sum_{n\neq m}
     \frac{f(E_n)-f(E_m)}
          {(E_n-E_m)^2}
      \label{sigmaxy1}
       \notag\\
     &\quad\times
     \mel{n}{\hat{v}_x}{m}
     \hspace{-1ex}
     \mel{m}{\hat{v}_y}{n},
     \\
       f(E) &= \frac{1}{e^{(E-\mu)/T}+1},
\end{align}
where $E_n$ and $\ket{n}$ are the $n$th eigenvalue and eigenvector of $H_{\mathrm{wp}}+H_{\mathrm{F}}$, respectively. 
The Fermi distribution function is denoted by $f(E)$ and $\hat{v}_i$ is velocity operator defined as
\begin{equation}
  \hat{v}_{i}\equiv \pdv{H}{(\hbar k_{i})}.
\end{equation}

\begin{figure}
    \centering
    \subfigure[$v_{2}^{2}>v_{s1}^{2}$]{
    \includegraphics[scale=0.25]{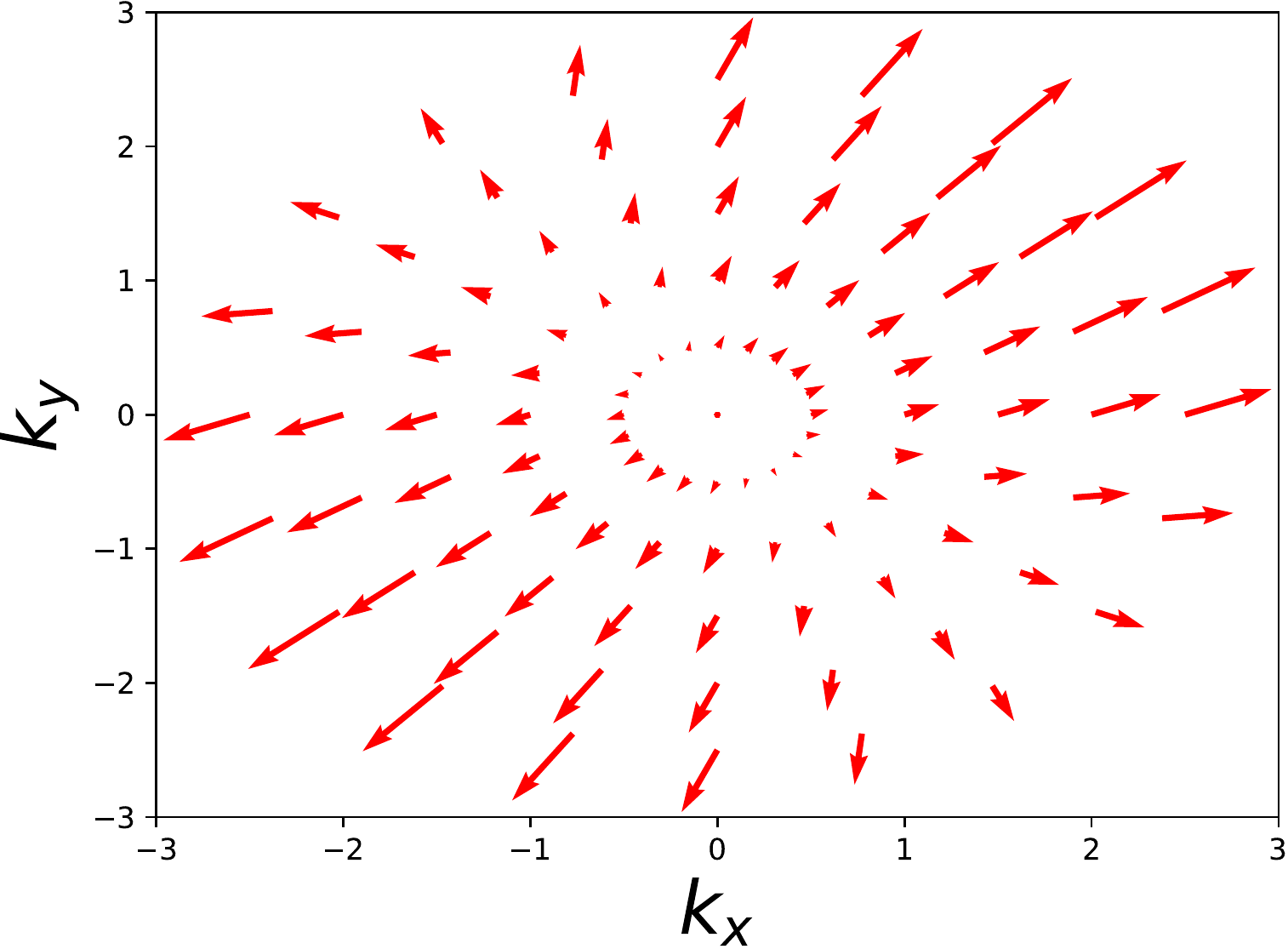}
    }
    \subfigure[$v_{2}^{2}<v_{s1}^{2}$]{
    \includegraphics[scale=0.25]{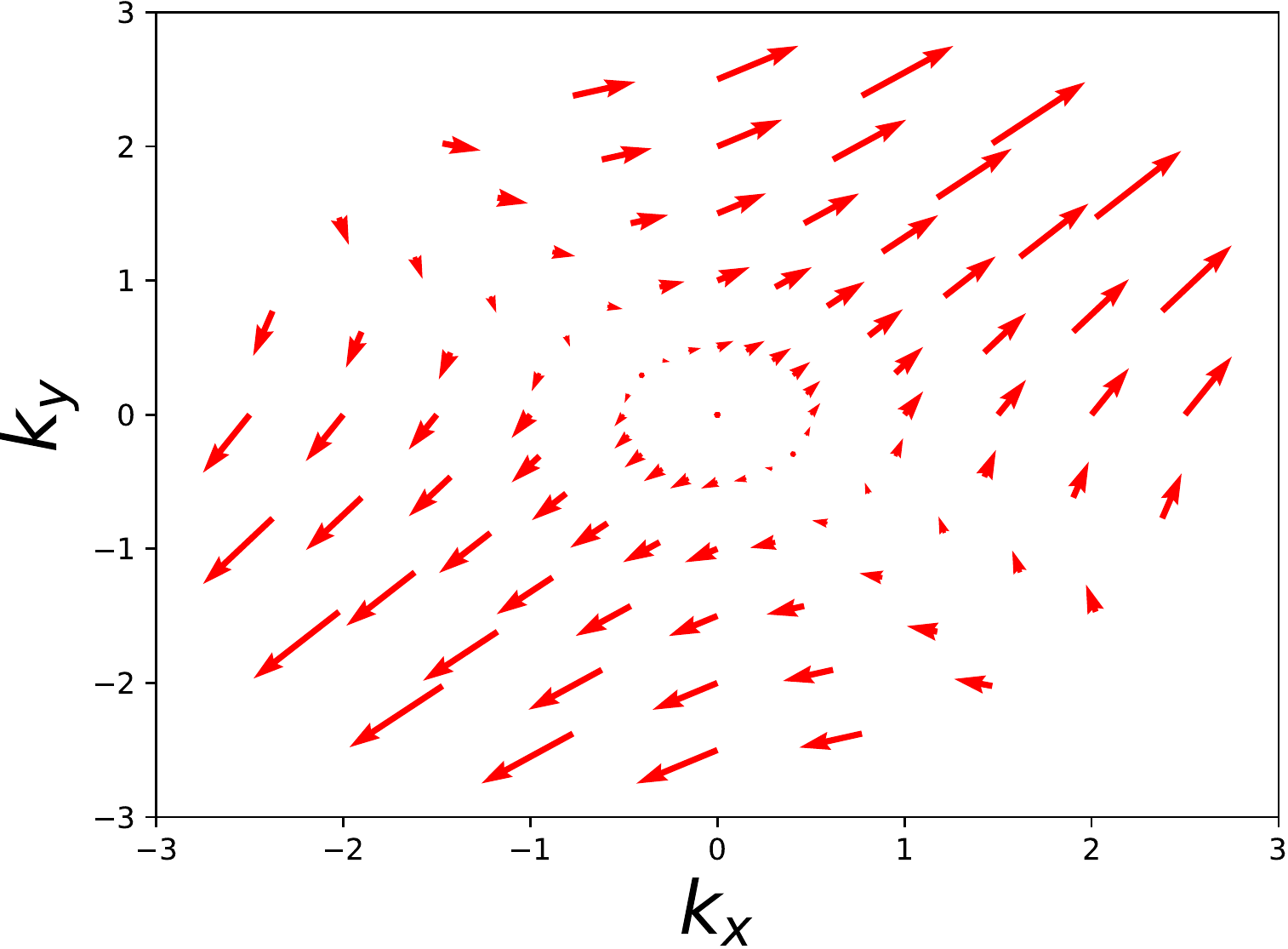}
    }
    \caption{Vector field $\vv{d}^+$ mapping onto $(k_x,k_y)$ plane.
    (a) Meron-like structure for $\tan^{-1}(v_{s1}/v_2)=\pi/8$.
    (b) Anti-meron-like structure for $\tan^{-1}(v_{s1}/v_2)=\pi/3$.}
    \label{fig_dvector}
\end{figure}

When the Fermi level is in the energy gap, the Hall conductivity is rewritten in terms of the Berry curvature of the occupied bands \cite{TKNN1982};
\begin{align}
    \si_{xy} 
    &= \frac{e^2}{h}\int\frac{d^2{k}}{2\pi}
    \sum_{X=\pm}
    \qty[\grad_{\vb*{k}} \times {\vb*{A}^X(\vb*{k})}]_z,
    \\
    \vb*{A}^X(\vb*{k}) 
    &= 
  -i \mel{\psi^X}{\grad_{\vb*{k}}}{\psi^X}.
\end{align}
For the calculation, we solve the eigenvalue problem of $H^{X}_{\mathrm{Fwp}} = H^{X}_{\mathrm{wp}} + H^{X}_{\mathrm{F}}$ as
\begin{align}
    H^{X}_{\mathrm{Fwp}}\ket{\psi^{X}} = sE_{X}\ket{\psi^{X}},
\end{align}
where $s=\pm 1$ and $E_X$ is given in Eq.~(\ref{Epm}).
The above equation is rewritten in the matrix form as
\begin{align}
	&
    \mqty(M-sE_{\pm} & \displaystyle\frac{v_{2}}{2}ke^{-i\theta_k}\pm i\frac{v_{s1}}{2}ke^{i\theta_k} \\
         \displaystyle\frac{v_{2}}{2}ke^{i\theta_k}\mp i\frac{v_{s1}}{2}ke^{-i\theta_k} & -M -sE_{\pm})
    \mqty(u^{\pm} \\ v^{\pm}) 
    \notag\\&
    =0,
\end{align}
with $k_x+ik_y=ke^{i\theta_k}$.
Therefore, we obtain the eigenstates as
\begin{align}
    \mqty(u^{\pm}\\v^{\pm}) 
    = \mqty(\sqrt{\displaystyle\frac{1}{2}\lb 1+\frac{sM}{E_{\pm}} \rb} \\
            se^{i\varphi_{\pm}} \sqrt{\displaystyle\frac{1}{2}\lb 1 - \frac{sM}{E_{\pm}} \rb})
    \equiv \ket{\psi^{\pm}}_{\rm I},
\end{align}
where $\varphi_{\pm}=\mathrm{arg}(\frac{v_{2}}{2}ke^{i\theta_k} \mp i\frac{v_{s1}}{2}ke^{-i\theta_k})$.
We define $\ket{\psi^{\pm}}_{\rm II}=e^{-i\varphi_{\pm}}\ket{\psi_{\pm}}_{\rm I}$. 
The Berry connection $\vb*{A}^X_{\rm I/II}(\vb*{k}) = -i \mel{\psi^X}{\grad_{\vb*{k}}}{\psi^X}_{\rm I/II}$ of valence band of $s=-1$ defined by $\ket{\psi^{\pm}}_{\rm I}$ ($\ket{\psi^{\pm}}_{\rm II}$) has the singularity at the origin ($k=0$) for $M>0$ ($M<0$).
Using the Stokes' theorem, therefore, the Hall conductivity is obtained by the contour integral for $k \to \infty$ as
\begin{align}
 \sigma_{xy}
 &
 = 
 \frac{e^2}{h}
 \int_{0}^{2\pi} \frac{d\theta_k}{2\pi}
 \sum_{X=\pm}
 \lim_{k \to \infty}
 \vb*{e}_{\theta_k}
\cdot
 \begin{cases}
 {\vb*{A}_{\rm II}^X(\vb*{k})},
 & M>0,
 \\
 \vb*{A}_{\rm I}^X(\vb*{k}),
 & M<0.
 \end{cases}
\end{align}
The Berry connection is asymptotically given by
\begin{align}
 \lim_{k \to \infty} \vb*{A}_{\rm I}^X(\vb*{k})
 &= \frac{1}{2} \pdv{\varphi_X}{\theta_k}
 \vb*{e}_{\theta_k},
 \\
 \lim_{k \to \infty} \vb*{A}_{\rm II}^X(\vb*{k})
 &= -\frac{1}{2} \pdv{\varphi_X}{\theta_k}
 \vb*{e}_{\theta_k}.
\end{align}
As a result, the Hall conductivity is obtained to be
\begin{align}
        \si_{xy} &=
         -\frac{e^{2}}{h}\sgn(M)\int^{2\pi}_{0}\frac{d\theta_k}{4\pi}
         \sum_{X = \pm}
         \pdv{\varphi_{X}}{\theta_k} 
        \notag\\
    &=-\frac{e^{2}}{h}\sgn(M)
    \notag\\& \quad \times
    \int^{2\pi}_{0}\frac{d\theta_k}{4\pi}
    \pdv{\theta_k} \lb-i\rb \ln(v_{2}^2e^{2i\theta_k}+v_{s1}^2e^{-2i\theta_k}) 
    \notag\\
    &=-\frac{e^2}{h}\sgn\qty(M)\sgn\qty(v_{2}^2-v_{s1}^{2}).
    \label{sigmaxy}
\end{align}
From this expression, one finds that the Hall conductivity changes its sign at $v_{s1}^2 = v_{2}^2$ without gap closing, which is not seen in the surfaces of topological insulators.
This change of sign is understood from the vector $\vv{d}^X$ defined as $H_{\rm Fwp}^X=\vv{d}^X\cdot\vv{\si}$.
The half-quantized Hall conductivity is equivalent to the winding number of the vector $\vb*{d}^X$  in the $k$ space.
For $v_{2}^{2}>v_{s1}^{2}$, the vector $\vv{d}^X$ has a meron-like structure shown in Fig.~\ref{fig_dvector}(a), resulting in the winding number $+1/2$.
On the other hand, for $v_{2}^{2}<v_{s1}^{2}$, the winding number is $-1/2$ because the vector $\vv{d}^X$ has an anti-meron-like structure shown in Fig.~\ref{fig_dvector}(b).
These two configurations are continuously connected since $\vb*{d}^X$ is a three-dimensional vector. 
Note that these configurations break the fourfold-rotation symmetry of the system.
In fact, for each sector the Hamiltonian $H_{\rm Fwp}^X$ contains the term $(v_{s1}/2) (k_x \sigma_y + k_y \sigma_x)$, which is fourfold-rotation odd, breaks the symmetry.
The total Hamiltonian restores the symmetry as $H_{\rm Fwp}^+$ and $H_{\rm Fwp}^-$ are swapped by the fourfold rotation, allowing the sign change of the quantized Hall conductivity without gap closing.  
This is a consequence of nonsymmorphic structure of glides.

\begin{figure}
  \centering
  \subfigure[$H_{\rm wp}^{1}$]{
    \includegraphics[scale=0.6]{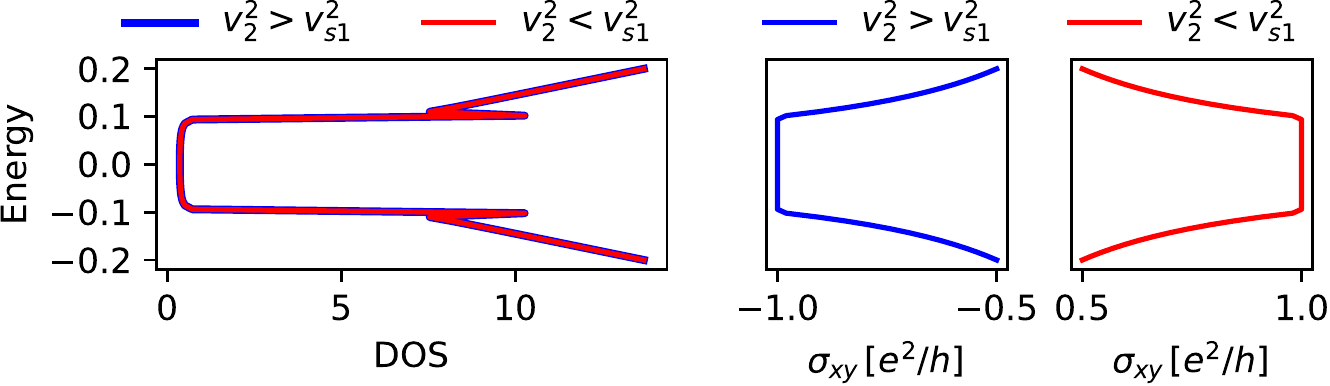}}
  \subfigure[$H_{\rm wp}^{1}+H_{\rm wp}^{2}$]{
    \includegraphics[scale=0.6]{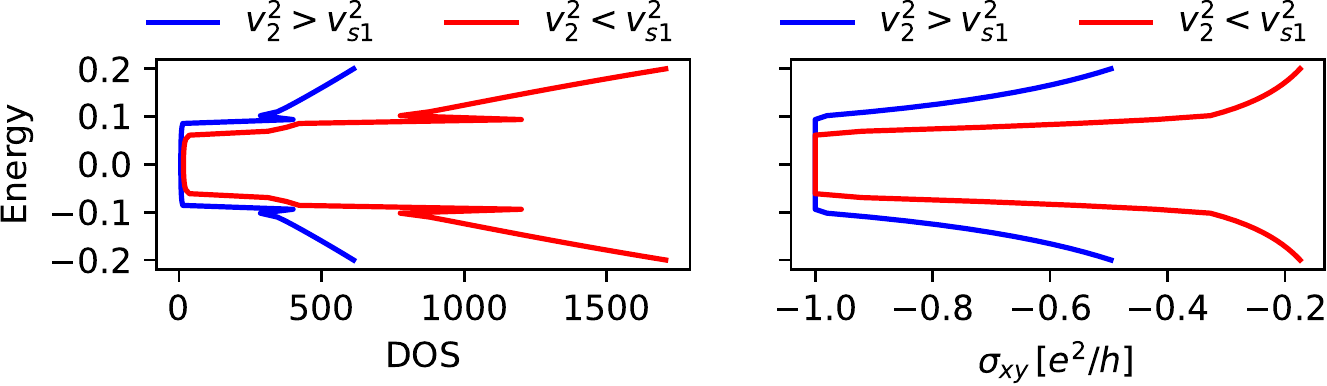}}
  \caption{Density of states (DOS) and the Hall conductivity for $M=0.1$ as a function of the Fermi energy. 
  (a) DOS and Hall conductivity in the linear model for $t_1=v_{\mathrm r1}=0, v_{s1}=0.5, v_{\mathrm s2}=0.8$ and $v_{\mathrm s2}^\p=0.6$ for the case of $v_2^2 > v_{\mathrm s1}^2$, while for $t_1=v_{\mathrm r1}=0, v_{\mathrm s1}=1.0, v_{\mathrm s2}=0.3$, and $v_{\mathrm s2}^\p=0.4$ for the case of $v_2^2 < v_{\mathrm s1}^2$. 
  (b) DOS and Hall conductivity in the model including the second-order terms for $v_{r1}=v_{s2}^\p=0.0,t_{1}=0.05, v_{s1}=0.3$, and $v_{s2}=0.2$ for the case of $v_{2}^2 > v_{\mathrm s1}^2$ while for $v_{r1}=v_{s2}^\p=0.0,t_{1}=0.05, v_{s1}=0.2$, and $v_{s2}=0.3$ for the case of $v_2^2 < v_{\mathrm s1}^2$. 
  }
  \label{Hall_calk}
\end{figure}

In addition to the analytical calculations limited to the first-order terms of the wavevector, we also numerically compute the Hall conductivity from Eq.~(\ref{sigmaxy1}).
Figure \ref{Hall_calk}(a) shows the verification of the plateau transition without gap closing discussed above. Within the energy gap ($|E|<0.1$), the Hall conductivity is quantized into $-e^2/h$ for $v_2^2 > v_{\mathrm s1}^2$ and $e^2/h$ for $v_2^2 < v_{s1}^2$, reproducing Eq.~(\ref{sigmaxy}).

This plateau transition does not occur in the system with the second-order terms.
Figure~\ref{Hall_calk}(b) shows the quantized Hall conductivity of $\sigma_{xy}=-e^2/h$ irrespective of the sign of $v_2^2-v_{\mathrm s1}^2$, implying that the plateau transition is an artifact of the model.  
If we consider a slab system with FM coupling instead of the effective surface model, the Hall conductivity is equivalent to the Chern number; hence no plateau transition without gap closing occurs. 
Therefore, linear terms alone are not sufficient for the minimal model of wallpaper fermion, and at least second-order terms must be included.

\subsection{Antiferromagnetic Wallpaper Fermion}
\label{AFM_WPF}
AFM moments induce an energy gap in the wallpaper fermion, which is a notable feature that distinguishes it from the case of a single Dirac fermion on the surface of a TI.
Here we focus on the SHC, which is nonzero for both the FM and AFM cases. 
The SHC is not quantized in the presence of spin-orbit interaction but partially characterizes the nontrivial topology of the surface state through the partial Chern number \cite{lin2022spin}.  
We show that in the strong coupling limit, the SHC remains finite for the AFM case while it vanishes for the FM case.

The Hamiltonian of the AFM coupling shown in Fig.~\ref{fig_magnet}(b) is written as
\begin{equation}
  H_{\rm AFM}=M\tau_z\si_z. 
  \label{Ham_AFM}
\end{equation}
%
\begin{figure}
  \centering
  \subfigure[3d plot]{
    \includegraphics[scale=0.4]{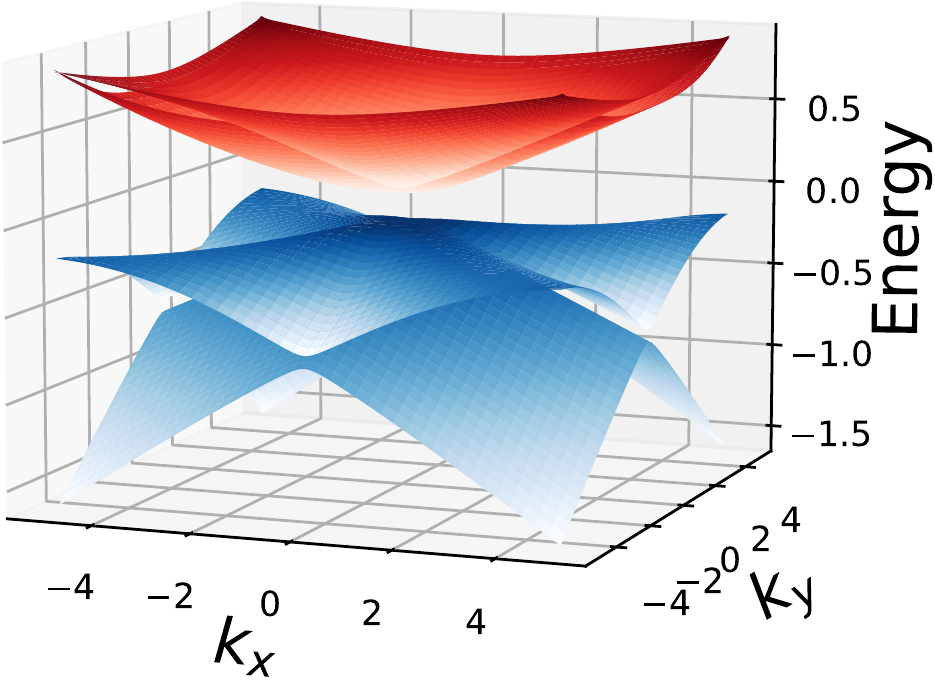}}
  \subfigure[2d plot]{
    \includegraphics[scale=0.65]{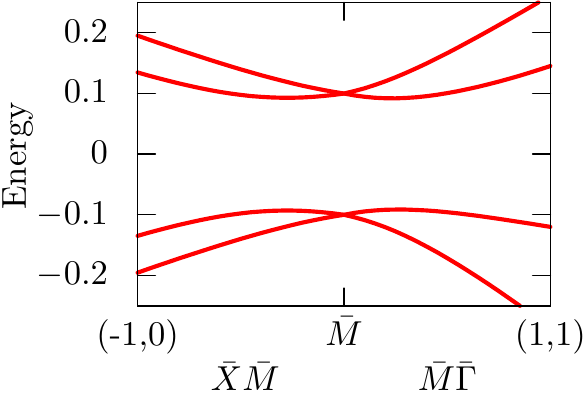}}
  \caption{
  Energy spectrum of the AFM wallpaper fermion for
  $t_1=0.03,
  v_{r1}=0.05,
  v_{s1}=0.1,
  v_{s2}=0.2,
  v_{s2}^{\p}=0.15$, and $M=0.1$.
  (a) The gap is induced by AFM coupling. 
  (b) Energy spectrum is not degenerate except for $k=0$ ($\bar M$ point). 
  }
  \label{Ene_antiferro}
\end{figure}
In Fig.~\ref{Ene_antiferro}, we show that the energy spectrum of the AFM wallpaper fermion $H_{\rm wp}+H_{\rm AFM}$ is gapped.

Next, we calculate the SHC.
The spin current operator is defined by
\begin{equation}
  \hat{J}^{z}_{i}\equiv \lB \hat{v}_i,\frac{\hbar \si_z}{2} \rB.
\end{equation}
Note that $\hat{J}^z_i = 0$ for the linear model. 
The SHC is written as 
\begin{align}
  \si^z_{xy} &= ie\hbar\int \frac{d^2k}{(2\pi)^2}\sum_{n\neq m}
               \qty[
                 f(E_n)-F(E_m)
               ]
 \notag\\ & \quad \times
               \frac{\bra{n} \hat{J}^z_x\ket{m}\bra{m} \hat{v}_y \ket{n}}
               {(E_n-E_m)^2} \label{siz_xy} ,\\
  J^{z}_{x} 
  &= \frac{1}{2}(
   t_{1}\tau_{x}\si_{z}+v_{r1}\tau_{y}
  )k_{y}.
\end{align}

\begin{figure}
  \centering
  \subfigure[]{
    \includegraphics[scale=0.65]{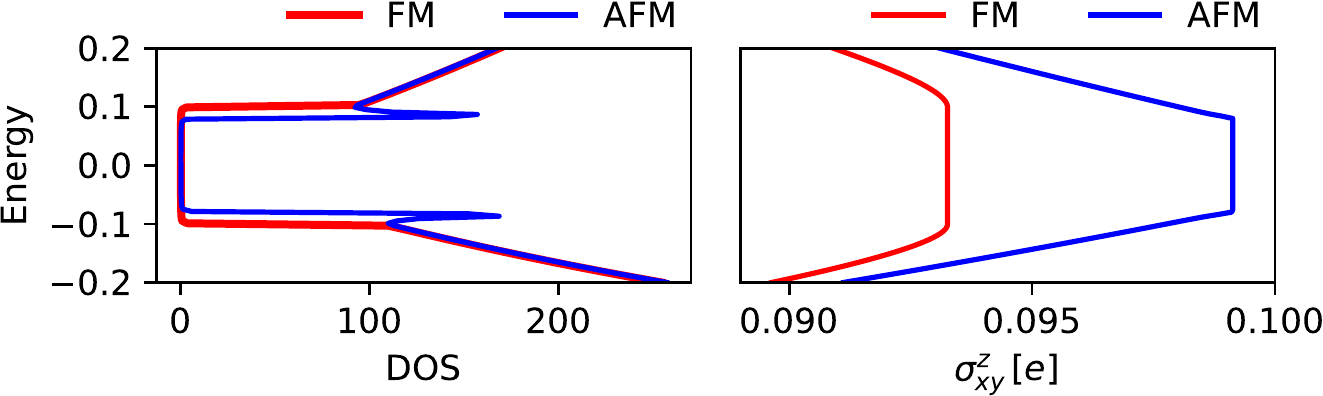}}
  \subfigure[]{
    \includegraphics[scale=0.65]{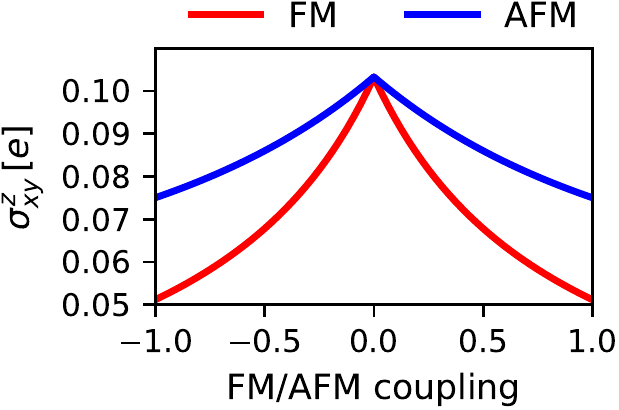}}
  \subfigure[]{
    \includegraphics[scale=0.65]{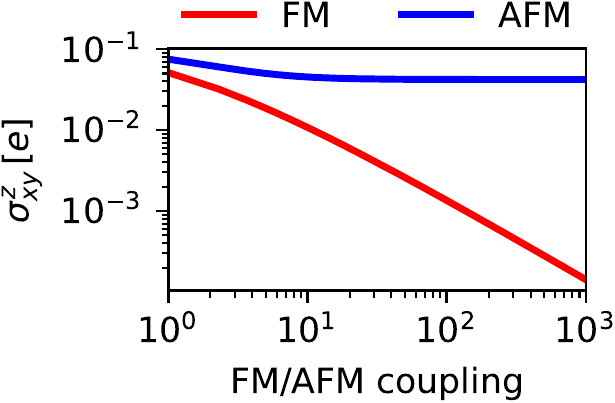}}
  \caption{
 Spin Hall conductivity as a function of the Fermi energy (a) and the FM/AFM coupling (b and c) for
  $t_1=0.1,
  v_{r1}=0.0,
  v_{s1}=0.3,
  v_{s2}=0.4,
  v_{s2}^{\p}=0.3$, and $M=0.1$.
  }
  \label{spin_c_fig}
\end{figure}

We show the SHC in Fig.~\ref{spin_c_fig}. 
As shown in Fig.~\ref{spin_c_fig}(a), the SHC has a plateau within the energy gap which takes a parameter-dependent value. 
The value of the plateau decreases as the FM/AFM coupling increases, as shown in Fig.~\ref{spin_c_fig}(b), reaching to be a finite value and zero in the strong coupling limit, respectively, as shown in Fig.~\ref{spin_c_fig}(c).
On the other hand, since the AFM case has a magnetic-reflection symmetry, the Hall conductivity is zero for the AFM case.

\subsection{Degeneracy of Energy Spectrum}

\begin{figure}
  \centering
  \subfigure[FM coupling]{
    \includegraphics[scale=0.7]{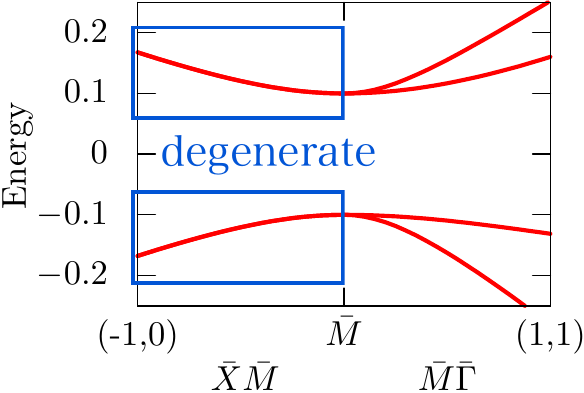}}
  \subfigure[AFM coupling]{
    \includegraphics[scale=0.7]{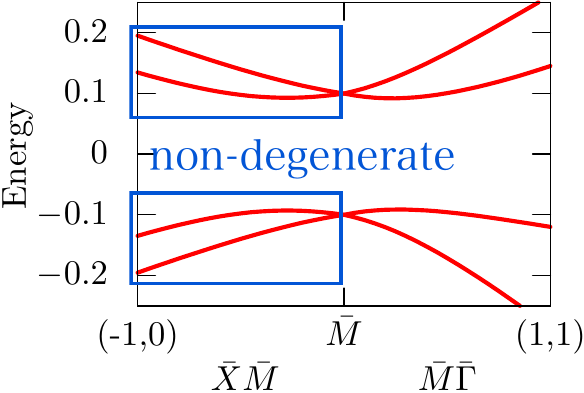}}
  \caption{
  Comparison of degeneracy on the $\bar{X}\bar{M}$ line for the FM (a) and AFM (b) cases. 
  }
  \label{comp_degene}
\end{figure}

We showed that the energy spectrum of the model with the FM (AFM) coupling, in which doubly degenerate bands (do not) appear along the $\bar X \bar M$ line, as reshown in Fig.~\ref{comp_degene}.
\begin{figure}
    \centering
    \includegraphics[scale=0.35]{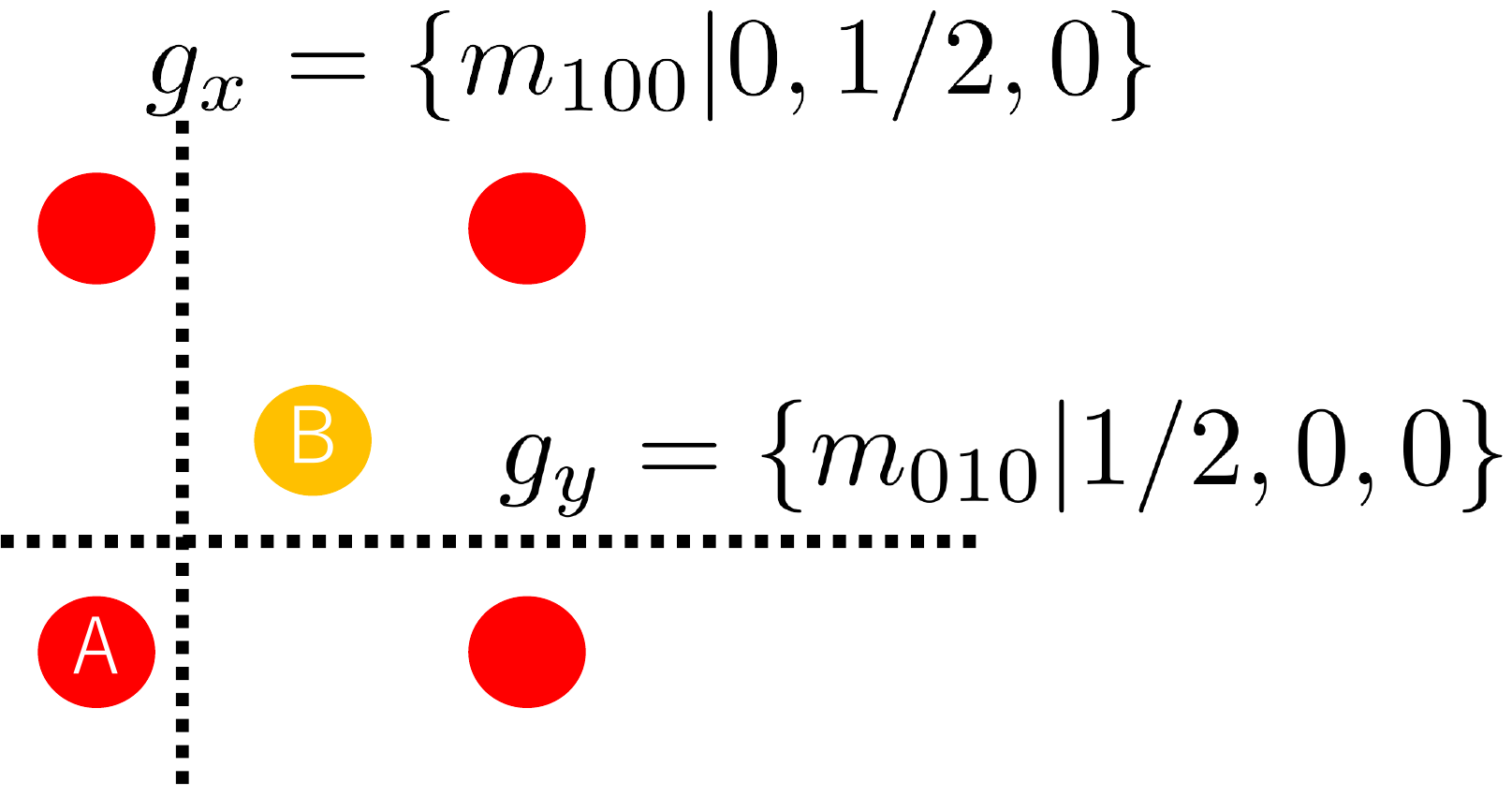}
    \caption{Schematic of glide symmetry of the model.
    The glides $g_x$ and $g_y$ flip the spins and exchange the sublattices A and B.}
    \label{glide}
\end{figure}
These behaviors are understood by TRS $\Theta$ and the glides $g_{x,y}$ the system holds, which are shown in Fig.~\ref{glide}.

In the nonmagnetic case, the degeneracy along the $\bar{X}\bar{M}$ line is protected by $\Theta g_y$ for $k_y=0$ and $\Theta g_x$ for $k_x=0$ \cite{wieder2018wallpaper}.
This symmetry is retained for the FM case and broken for the AFM case because the glide exchanges the sublattices A and B and reverses the $z$ component of spin, and TRS $\Theta$ also reverses the $z$ component of spin but does not exchange the sublattices.
Therefore, the FM case shows the degeneracy on the $\bar X \bar M$ line by the symmetry $\Theta g_x$ and $\Theta g_y$ while the AFM case does not. 

\section{Symmetry--adapted model}
\label{more_G_framework}

So far, we have worked on a concrete lattice model. 
In this section, we stress that the results from the model are irrelevant to the details of the system, constructing a general effective Hamiltonian from symmetry consideration. 

We construct an effective model for a wallpaper fermion on the $(001)$ surface of a crystal with the space-group $P4bm$ (No.~100) symmetry. 
The surfaces where a wallpaper fermion emerges have either $pgg$ or $p4g$ wallpaper-group symmetry. We concentrate on $p4g$, which corresponds to the $P4bm$ space group without $\ev{001}$ translations. This implies that analyzing the irreducible representation of $\bar M$ point in the $P4bm$ space group is adequate.
The wallpaper fermion is realized on the $\bar M$ point as a four-dimensional irreducible representation (irrep) that is the $\bar M_6 \bar M_7$ irrep \cite{Aroyo2011-kj, Aroyo2006-bn, Aroyo2006-jv, Elcoro2021-gs, Xu2020-lo}.
The $\bar M_6 \bar M_7$ irrep of the generators is given by
\begin{align}
 D(\{4^+_{001} | 0,0,0\})
 &= -\frac{1}{\sqrt 2} 
 \sigma_0 \tau_z 
 + i \frac{1}{\sqrt 2} \sigma_z \tau_0,
 \\
 D(\{m_{010} | 1/2, 1/2, 0\})
 &=
 -\frac{1}{\sqrt 2} \sigma_0 \tau_x
 - \frac{1}{\sqrt 2} \sigma_z \tau_y,
\end{align}
and time reversal
\begin{equation}
	\Theta=-i\si_y\tau_0K.
\end{equation}
By using these representations, we take the irreducible decomposition of 16 matrices $\sigma_\mu \tau_\nu$, $\mu, \nu = 0, 1, 2, 3$, which is summarized in Table \ref{table_100_sym}.

\begin{table*}
\begin{ruledtabular}
\caption{Irreducible decomposition of matrices on the $M$ point of $P4bm$.}
\label{table_100_sym}
 \begin{tabular}{lllllll}
      &  $E$ & $2C_4$ & $2c_v$ & basis & TR-even & TR-odd
      \\
      \hline
   $A_1$   & 1 & 1 & 1 & $z$ & $\sigma_0 \tau_0$ & $\sigma_z \tau_0$ 
   \\
   $A_2$ & 1 & 1 & $-1$ & $xy(x^2-y^2)$ & $\sigma_0 \tau_z$ & $\sigma_z \tau_z$
   \\
   $B_1$ & 1 & $-1$ & 1 & $x^2-y^2$ & & $\sigma_y \tau_x$, $\sigma_x \tau_x$
   \\
   $B_2$ & 1 & $-1$ & $-1$ & $xy$ & $\sigma_x \tau_y$, $\sigma_y \tau_y$
   \\
   $E$ & 2 & 0 & 0 & $(x,y)$ & $(\si_0 \tau_x + \si_z \tau_y, \si_0 \tau_x - \si_z \tau_y)$ & \begin{tabular}{c}
        $(\si_z \tau_x + \si_0 \tau_y, \si_z \tau_x - \si_0 \tau_y )$ \\
        $(\si_y \tau_z + \si_x \tau_0, \si_y \tau_z - \si_x \tau_0)$ \\
        $(\si_x \tau_z - \si_y \tau_0, \si_x \tau_z + \si_y \tau_0)$
   \end{tabular}
 \end{tabular}
 \end{ruledtabular}
\end{table*}

The effective Hamiltonian is given as a totally symmetric representation. 
Two-dimensional momentum $(k_x, k_y)$ belongs to the TR-odd $E$ irrep and is coupled with the same irreps $(\sigma_z \tau_x + \sigma_0 \tau_y, \sigma_z \tau_x - \sigma_0 \tau_y)$, $(\sigma_y \tau_z + \sigma_x \tau_0, \sigma_y \tau_z - \sigma_x \tau_0)$, and $(\sigma_x \tau_z - \sigma_y \tau_0, \sigma_x \tau_z + \sigma_y \tau_0)$ as
\begin{equation}
  \begin{split}
    \frac{k_x}{\sqrt{2}}
      (\si_z \tau_x + \si_0 \tau_y)
    +\frac{k_y}{\sqrt{2}}
      (\si_z \tau_x - \si_0 \tau_y)
     \equiv k_x\al_1+k_y\al_2 ,
  \end{split}
\end{equation}
\begin{equation}
  \begin{split}
    \frac{k_x}{\sqrt{2}}
      (\si_y \tau_z + \si_x \tau_0)
      +\frac{k_y}{\sqrt{2}}
      (\si_y \tau_z - \si_x \tau_0)
      \equiv k_x\al_3+k_y\al_4 ,
  \end{split}
\end{equation}
and
\begin{equation}
  \begin{split}
    \frac{k_x}{\sqrt{2}}
      (\si_x \tau_z - \si_y \tau_0)
    +\frac{k_y}{\sqrt{2}}
      (\si_x \tau_z + \si_y \tau_0)
     \equiv k_x\al_5+k_y\al_6 ,
  \end{split}
\end{equation}
resulting in the Hamiltonian
\begin{equation}
  \begin{split}
    H_{\rm Gwp}
      =v_{1}(k_x\al_1+k_y\al_2)
       &+v_{2}(k_x\al_3+k_y\al_4)\\
       &+v_{3}(k_x\al_5+k_y\al_6) ,
    \label{Ham_Gwp}
  \end{split}
\end{equation}
where $v_i$ is an arbitrary real number.
In this basis, the $z$ component of the FM moment, which belongs to the TR-odd $A_2$ irrep, is represented by $M \tau_z \si_z$ and the $z$ component of the AFM moment, which belongs to the TR-odd $A_1$ irrep, by $M\tau_0 \si_z$. 
On the other hand, the $x$ and $y$ components of the moments cannot be uniquely determined 
only by the symmetry consideration.

\begin{figure}
  \centering
  \subfigure[Energy spectrum for the nonmagnetic case around the $\bar M$ point]{
    \includegraphics[scale=0.4]{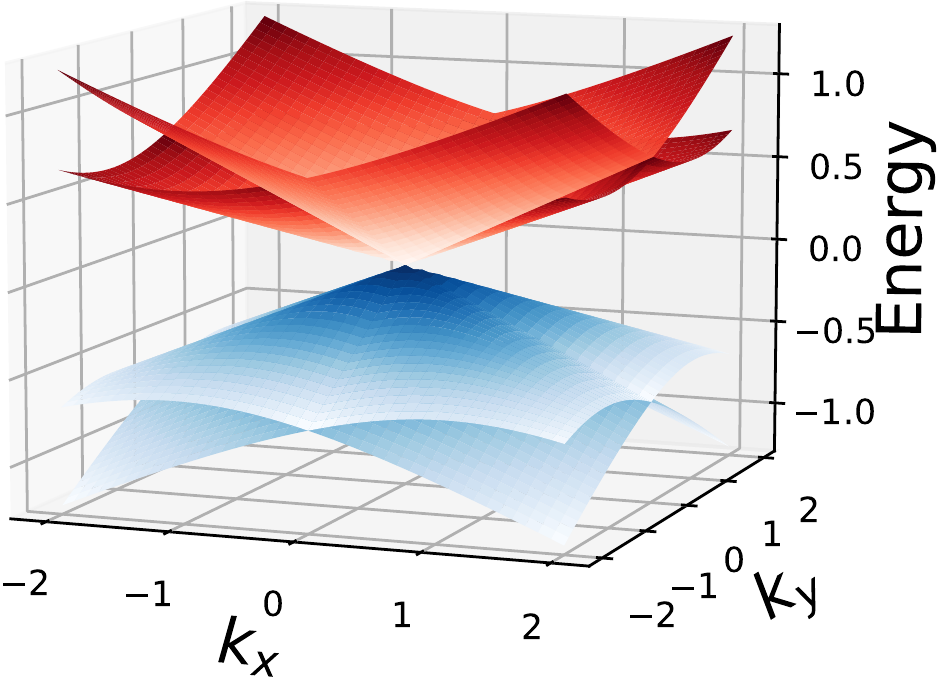}}
  \subfigure[Energy spectrum along the $\bar X \bar M$ and $\bar M \bar \Gamma$ lines]{
    \includegraphics[scale=0.65]{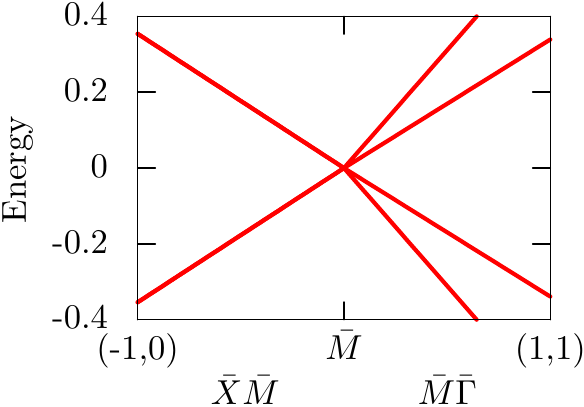}}
  \subfigure[FM case]{
    \includegraphics[scale=0.65]{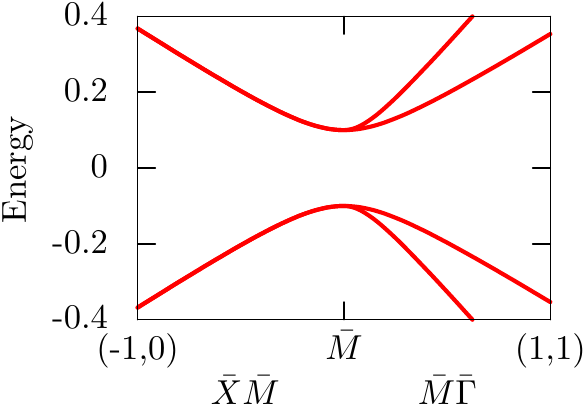}}
  \subfigure[AFM case]{
    \includegraphics[scale=0.65]{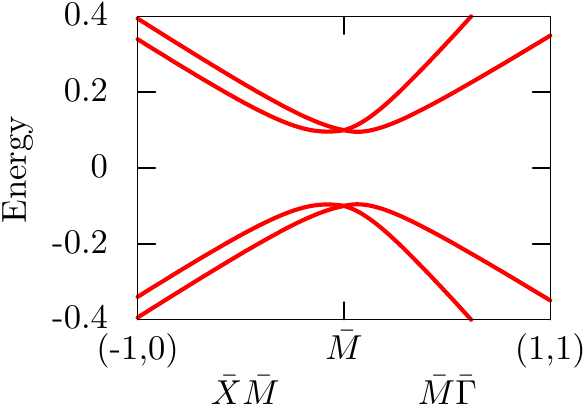}}
  \caption{
  Eigenvalues of the Hamiltonian
  for the generalized model (\ref{Ham_Gwp}) with 
  $
  v_{1}=0.1,
  v_{2}=-0.23,
  v_{3}=0.25$, and $M=0.1$.
  (a and b) Nonmagnetic case. (c and d) FM and AFM cases.
  (a) Energy spectrum around the $\bar M$ point.
  (b) There exist twofold degeneracy on the $\bar X \bar M$ line and fourfold degeneracy at the origin.
  (c) The energy spectrum is gapped by the FM coupling ($M\tau_z \si_z$). 
  Twofold degeneracy on the line $k_y=0$ remains.
  (d) The energy spectrum is gapped by the AFM coupling ($M\tau_0 \si_z$). 
  There is no degeneracy except at the origin. 
  }
  \label{Ene_general}
\end{figure}

We verify that the Hamiltonian (\ref{Ham_Gwp}) reproduces those of Eq.~(\ref{Hsurface}).
The nonmagnetic case [Fig.~\ref{Ene_general}(a) and \ref{Ene_general}(b)] has hosts the fourfold degeneracy on the $\bar M$ point ($k=0$), while the FM [Fig.~\ref{Ene_general}(c)] and the AFM [Fig.~\ref{Ene_general}(d)] cases host a gapped spectrum with and without degenerate bands on the $\bar X \bar M$ line, respectively. 
Therefore, the Hamiltonian (\ref{Ham_Gwp}) reproduces that for the wallpaper fermion.
Note that the model derived here is generic since it is constructed solely based on the symmetry of the system. 

The second-order terms of momentum is also derived in the same manner. 
$(k_{x}^2-k_{y}^2)$ is the TR-even $B_1$ irrep and does not appear in the Hamiltonian due to the absence of matrix of the same irrep, while $k_{x}k_{y}$ of the TR-even $B_2$ irrep is coupled with $\sigma_x \tau_y$ and $\sigma_y \tau_y$. 

Finally, we show that the Hamiltonians (\ref{Ham_Gwp}) and (\ref{Hsurface}) are unitary equivalent. 
We set the unitary matrix $U$ as
\begin{equation}
    U = \mqty(e^{i7\pi/8} & 0 & 0 & 0 \\
              0 & 0 & 0 &  e^{i\pi/8} \\
              0 & 0 &  e^{i\pi/8} & 0 \\
              0 & e^{-i\pi/8} & 0 & 0 \\
         ),
\end{equation}
in the basis that the matrix $\sigma_\mu \tau_\nu$ is represented as
\begin{align}
 \sigma_0 \tau_\nu
 = \pmqty{
   \tau_\nu & 0
   \\
   0 & \tau_\nu
 },
 \
  \sigma_1 \tau_\nu
 = \pmqty{
 	0 & \tau_\nu
 	\\
 	\tau_\nu & 0
 },
 \\
  \sigma_2 \tau_\nu
 = \pmqty{
 	0 & -i \tau_\nu
 	\\
 	i\tau_\nu & 0
 },
 \
 \sigma_3 \tau_\nu
 = \pmqty{
 	\tau_\nu & 0
 	\\
 	0 & -\tau_\nu
 }.
\end{align}
The matrices are transformed by $U$ as
\begin{align}
    &U^{\dag}\sigma_{z}\tau_{z}U = +\sigma_{z}\tau_{0}, & U^{\dag}\sigma_{z}\tau_{0}U = +\sigma_{z}\tau_{z}, \\
    &U^{\dag}\al_{1}U = -\sigma_{x}\tau_{x} , &U^{\dag}\al_{2}U = +\sigma_{y}\tau_{x}, \\
    &U^{\dag}\al_{3}U = -\sigma_{x}\tau_{z} , &U^{\dag}\al_{4}U = -\sigma_{y}\tau_{z}, \\
    &U^{\dag}\al_{5}U = +\sigma_{y}\tau_{0} , &U^{\dag}\al_{6}U = -\sigma_{x}\tau_{0}, \\
    &U^{\dag}\sigma_{x}\tau_{y}U = -\sigma_{z}\tau_{y}, &U^{\dag}\sigma_{y}\tau_{y}U = +\sigma_{0}\tau_{x}. \label{gene_toy_secnd_ord}
\end{align}
Thus, the Hamiltonian from symmetry consideration is obviously equivalent to the Hamiltonian for the four-sublattice model.

\section{Discussion}
\label{Discuss}
We note the scope of our effective model. 
The FM and AFM couplings expressed in Eqs.~(\ref{Ham_FM}) and (\ref{Ham_AFM}) are assumed to be spatially uniform. 
This situation can be realized when the system intrinsically becomes the FM or AFM insulating phase.  
On the other hand, our theory can apply to the junction systems attached to an FM or AFM insulators, where the coupling constant is estimated as $\sum_{i= -\infty}^0 \psi_i^\dag M_i \psi_i$ with the spatially decaying function $M_i$ for proximity-induced magnetization.

On the other hand, we assume that the FM/AFM moments are small compared to the band gap in the bulk.
If the moments are substantially large, the system goes into a topologically trivial or different topological state, e.g., a gapless wallpaper fermion in magnets \cite{hwang2022magnetic}, which is beyond the scope of the present study.

\section{Summary}
\label{summary}
In this study, we found that wallpaper fermion can couple with both FM and AFM moments and
result in the Hall effects.
Firstly, when the FM coupling is present, the Hall conductivity is quantized into twice the value predicted for TIs, contributed equally from double Dirac cones of the wallpaper fermion.
Additionally, the sign of Hall conductivity for the linear-order model, superposition of independent double Dirac cones, can be inverted without closing the energy gap. 
This anomalous behavior is due to the fact that each Dirac cone breaks the fourfold rotational symmetry of the system, while the entire wallpaper fermion remains symmetric.
The second-order terms hybridize the Dirac cones, resolving this anomaly.
Combined with the fact that such an anomaly does not occur in the slab system, the linear-order model is implied to be invalid for wallpaper fermions.
Secondly, SHC is suppressed by both FM and AFM coupling in a different way from each other. 
The SHC goes to zero in the strong coupling limit for the FM case, while remains finite for the AFM case. 
Thirdly, the eigenvalues of Hamiltonian are degenerate on the $\bar{X}\bar{M}$ line for the FM case, while this degenerate is lifted by the AFM coupling breaking the magnetic glide symmetry. 
Finally, the effective Hamiltonian was proved valid for general wallpaper fermions by deriving the symmetry-adapted form of the Hamiltonian. 
These results provide a basis for clarifying the transport phenomena, including spintronics, using wallpaper fermions. 

\bibliography{afmwf.bib}
\end{document}